\documentclass[trackchanges,twocolumn]{aastex701}

\begin{document}

\title{Testing a Pre-Supernova Contribution to $^{44}$Ti in Cassiopeia A}

\author[orcid=0000-0001-9267-1693]{Toshiki Sato}
\affiliation{Department of Physics, School of Science and Technology, Meiji University, Kanagawa, 214-8571, Japan}
\email[show]{toshiki@meiji.ac.jp}  

\author[orcid=0000-0002-1283-6636]{Joshua Issa}
\affiliation{Astronomy Research Centre, Department of Physics \& Astronomy, University of Victoria, V8P 5C2, Victoria, Canada}
\affiliation{NuGrid Collaboration}
\email[]{joshuaissa@uvic.ca}

\author[orcid=0000-0001-8087-9278]{Falk Herwig}
\affiliation{Astronomy Research Centre, Department of Physics \& Astronomy, University of Victoria, V8P 5C2, Victoria, Canada}
\affiliation{NuGrid Collaboration}
\email[]{fherwig@uvic.ca}

\author[orcid=0009-0000-8310-9463]{Yui Kuboike}
\affiliation{Department of Physics, School of Science and Technology, Meiji University, Kanagawa, 214-8571, Japan}
\email[]{}

\author[orcid=0000-0002-3362-7040]{J. Martin Laming}
\affiliation{Space Science Division, Code 7684, Naval Research Laboratory, Washington DC 20375, USA}
\email[]{}

\author[orcid=0009-0003-0653-2913]{Kai Matsunaga}
\affiliation{Department of Physics, Kyoto University, Kyoto 606-8502, Japan}
\email[]{}

\author[orcid=0000-0003-4876-5996]{Ryo Sawada}
\affiliation{Institute for Cosmic Ray Research, The University of Tokyo, Chiba, 277-8582, Japan}
\affiliation{Department of Earth Science and Astronomy, The University of Tokyo, Tokyo, 153-8902, Japan}
\email[]{}

\begin{abstract}
The radioactive isotope $^{44}$Ti in Cassiopeia A has been used as a tracer of the innermost ejecta and explosion asymmetries in core-collapse supernova explosions. However, recent stellar-evolution models suggest substantial $^{44}$Ti may also be synthesized prior to explosion during oxygen–carbon (O–C) shell mergers, a process proposed for the Cas A progenitor. We investigate this pre-supernova contribution by combining NuSTAR measurements of the spatial and kinematic distribution of $^{44}$Ti with X-ray and infrared ejecta diagnostics. The low- and intermediate-velocity $^{44}$Ti components are more closely associated with O-layer tracers than with shocked Fe-rich material. If the $^{44}$Ti associated with the O-layer tracers was synthesized prior to core collapse, it could contribute up to \(\sim7\times10^{-5}\,M_\odot\), about half of the total. We find that this amount of $^{44}$Ti can be produced in O–C shell-merger models with ingestion rates and convective velocities higher than those in standard one-dimensional stellar-evolution calculations. Stable Fe-group abundance ratios measured with XRISM provide an additional constraint on the explosive contribution, as three-dimensional models with larger $^{44}$Ti yields tend to predict Ti/Fe and Mn/Cr ratios above the observed values. These results suggest that both pre-supernova shell-merger nucleosynthesis and explosive burning contribute to the observed $^{44}$Ti.
\end{abstract}

\section{Introduction} 

The young supernova remnant Cassiopeia A (Cas A) provides a unique laboratory for studying nucleosynthesis in core-collapse supernovae \citep[e.g.,][]{2000ApJ...528L.109H,2012ApJ...746..130H,2020ApJ...893...49S,2023ApJ...954..112S,2021Natur.592..537S,2022ApJ...932...93T,2025ApJ...990..103S}. Among the most important diagnostics is the radioactive isotope $^{44}$Ti, whose decay lines at 67.9 and 78.4 keV have been detected and spatially resolved by NuSTAR \citep[e.g.,][]{2014Natur.506..339G,2017ApJ...834...19G}. These observations have revealed a highly asymmetric three-dimensional distribution of $^{44}$Ti, offering direct insight into the innermost ejecta \citep[see also][for other $^{44}$Ti observations from Cas A]{1998PASP..110..637D,2001ApJ...560L..79V,2006ApJ...647L..41R,2009A&A...502..131M,2015A&A...579A.124S}.

Traditionally, $^{44}$Ti in Cas A has been interpreted as a product of explosive silicon (Si) burning under $\alpha$-rich freeze-out conditions \citep[e.g.,][]{2006ApJ...640..891Y,2010ApJS..191...66M,2017ApJ...842...13W,2020ApJ...895...82V,2020A&A...638A..83W,2023ApJ...957L..25S,2024ApJ...974...39W,2024ApJ...962...71W}. In this framework, the spatial relationship between $^{44}$Ti and Fe-group elements is expected to reflect the geometry and dynamics of the explosion. However, some observational features challenge a simple one-to-one correspondence between $^{44}$Ti and Fe, including the weak spatial correlation between them \citep{2014Natur.506..339G,2017ApJ...834...19G}. Recent long-term 3D neutrino-driven simulations suggest that a simple one-to-one mapping between $^{44}$Ti and Fe-group material is not generally expected \citep{2023ApJ...957L..25S,2024ApJ...974...39W,2024ApJ...962...71W}. In these models, the thermodynamic and $Y_e$ histories of individual ejecta parcels are highly heterogeneous (and partly stochastic), producing substantial spatial and model-to-model variations in the local $^{44}$Ti/Fe (or $^{44}$Ti/$^{56}$Ni) yield ratio. Moreover, $^{44}$Ti synthesis can continue for a longer time than $^{56}$Ni synthesis, allowing the two isotopes to be produced at different times and locations. This difference may help explain the weak spatial correspondence between $^{44}$Ti and Fe seen in the NuSTAR maps. 

Recent theoretical studies have proposed another possible origin of $^{44}$Ti, in which it is synthesized prior to explosion during late-stage stellar evolution in massive stars. \citet{2017ApJ...836...79C} showed that stellar rotation can lead to the formation of extended convective O-burning shells in which $^{44}$Ti is produced during stellar evolution and may survive the subsequent explosion, implying that a pre-supernova contribution to the total $^{44}$Ti yield cannot be excluded under certain conditions. More recently, \citet{2026ApJ..1004...52I} performed detailed one-dimensional (1D) calculations of an O shell during a late-stage oxygen-carbon (O-C) shell merger and demonstrated that such mergers can produce substantial amounts of $^{44}$Ti prior to core collapse given mixing conditions motivated by three-dimensional (3D) hydrodynamic simulations \citep[see also][]{2026ApJ..1003L...2B}. They further showed that the resulting pre-supernova $^{44}$Ti yield is highly sensitive to the degree of mixing and thermodynamic conditions, and in some cases can be comparable to or even exceed the explosive contribution. These results motivate an observational examination of whether signatures of such pre-supernova $^{44}$Ti production can be identified in supernova remnants such as Cas A.

Independently of these theoretical predictions regarding the synthesis of $^{44}$Ti, observational evidence has been presented that the progenitor of Cassiopeia A experienced an O–C shell merger shortly before core collapse \citep{2025ApJ...990..103S,2026NatAs..10..144X}. Based on spatially resolved abundance measurements of O-rich ejecta, \citet{2025ApJ...990..103S} showed that systematic variations in the Ne/Mg and Si/Mg ratios require incomplete mixing between adjacent burning shells, providing direct evidence for merger-driven convection in the late stages of stellar evolution. In addition, recent XRISM observations revealed significant enrichment of odd-Z elements such as Cl and K in Cas A, which cannot be explained by standard supernova explosion models \citep{2026NatAs..10..144X}. The enhanced abundances of these odd-Z elements, spatially associated with O-rich ejecta, instead point to non-standard stellar evolutionary processes, including shell mergers \citep[e.g.,][]{2018MNRAS.474L...1R,2026ApJ...997..314I}.

\cite{2026ApJ..1003L...2B} provided an important theoretical framework for assessing the contribution of O–C shell mergers to the observed \(^{44}\mathrm{Ti}\) in Cas A. Using extensive grids of one-dimensional progenitor models and spherically symmetric explosion calculations, they showed that ejecta-integrated abundance ratios, particularly Ar/Ne, can be used as diagnostics of a shell merger. For models compatible with Cas A, they then estimated the merger-produced \(^{44}\mathrm{Ti}\) yield and examined whether this material, if distributed over the outer ejecta, could remain below the NuSTAR detection threshold in individual spatial regions. This analysis connects model predictions for the merger contribution with the observational sensitivity to an extended outer component. At the same time, the production of \(^{44}\mathrm{Ti}\) in a convective-reactive shell merger is expected to depend sensitively on ingestion and mixing processes that are not fully represented in one-dimensional stellar-evolution calculations \citep{2026ApJ..1004...52I}. Moreover, because their observational test concerns a potentially undetected outer component, it leaves open whether the \(^{44}\mathrm{Ti}\) actually detected in Cas A is associated with ejecta originating from the oxygen layer. These considerations motivate further investigations from both observational and theoretical perspectives.

If part of the observed \(^{44}\mathrm{Ti}\) was synthesized in the oxygen-burning layer during a pre-supernova O–C shell merger, that component should retain a spatial and kinematic association with ejecta carrying O-layer nucleosynthetic signatures. Subsequent multidimensional mixing and shock interaction may prevent a one-to-one correspondence, but some association with O-rich ejecta and related tracers such as Si and Ar is nevertheless expected. Here, we test this prediction by comparing the detected \(^{44}\mathrm{Ti}\) distribution with shocked and unshocked O-layer tracers observed with Chandra and Spitzer, and use these associations to place an empirical upper bound on the possible pre-supernova contribution. We then compare this observational constraint with shell-merger calculations that explore a range of ingestion rates and convective mixing efficiencies motivated by multidimensional hydrodynamic simulations. As an independent test of the explosive contribution, we also compare stable Fe-group abundances measured with XRISM with predictions from multidimensional supernova models. 

\begin{figure*}[t]
 \begin{center}
  \includegraphics[bb=0 0 2250 975,width=18cm]{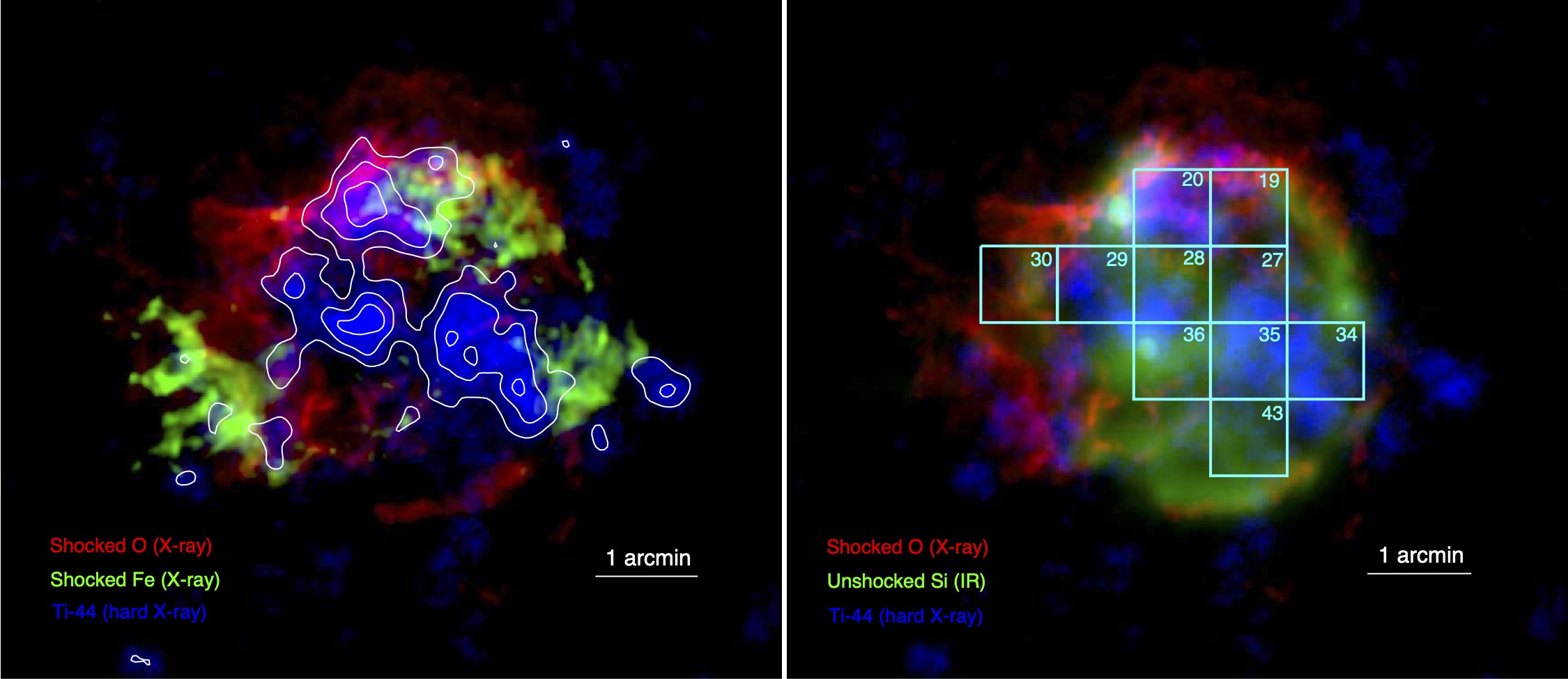}
 \end{center}
\caption{{\it Left}: The spatial distribution of $^{44}$Ti in Cas A (blue) compared with the other bright X-ray features (red: oxygen, green: iron). The oxygen traces material from the oxygen layer formed during stellar evolution. The white contours represent the intensity distribution of $^{44}$Ti. {\it Right}: Same as the left figure, but compared with the infrared image of unshocked Si (green) obtained with Spitzer \citep[e.g.,][]{2008ApJ...673..271R,2009ApJ...693..713S}. Cyan boxes with associated region numbers indicate where $^{44}$Ti has been detected in \cite{2017ApJ...834...19G}.}
\label{fig:f1}
\end{figure*}

\begin{figure*}[t]
 \begin{center}
  \includegraphics[bb=0 0 2147 1072,width=18cm]{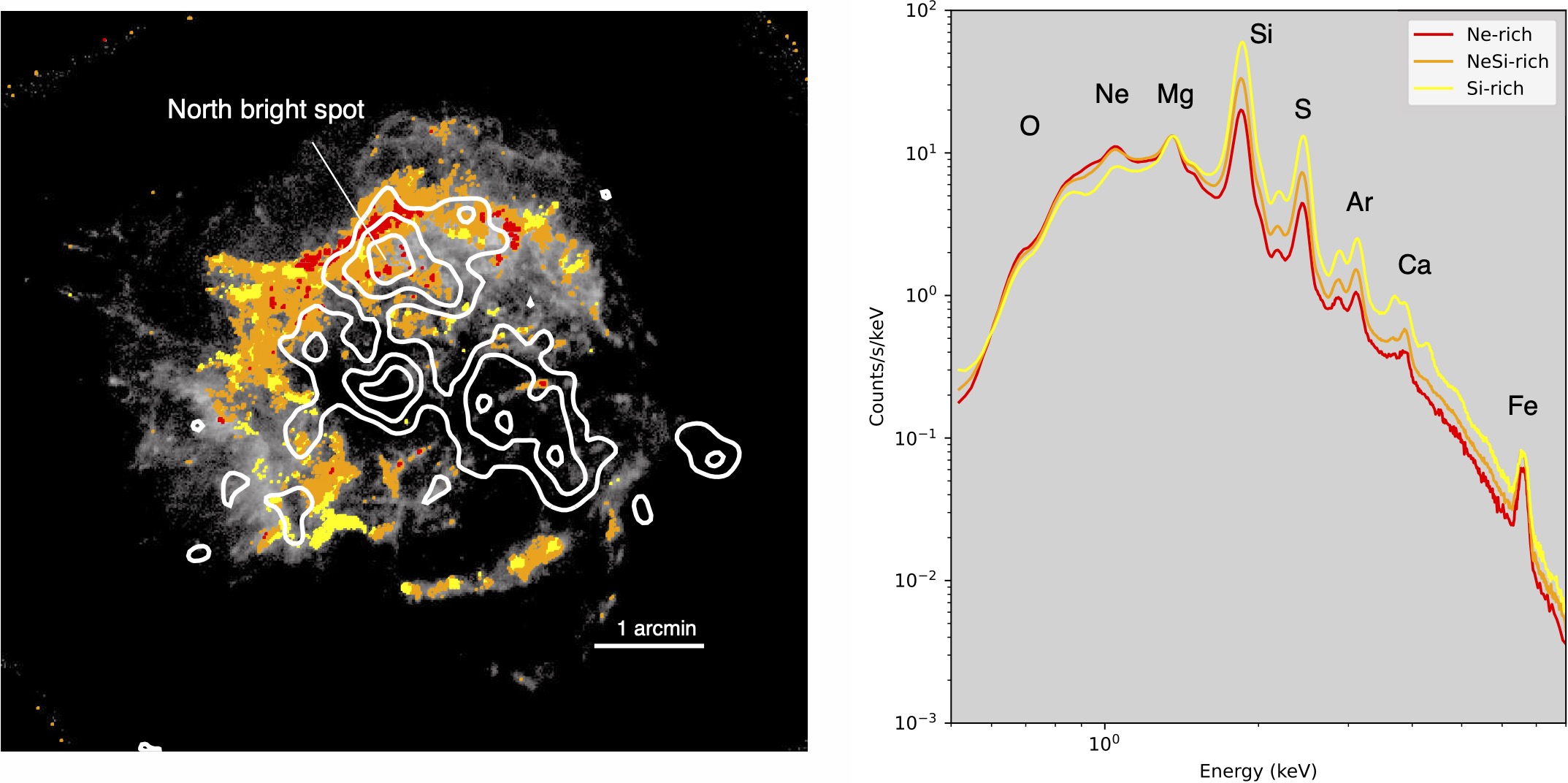}
 \end{center}
\caption{{\it Left}: Spatial classification of oxygen-rich ejecta in Cassiopeia A based on a principal component analysis of narrow-band X-ray images (Y. Kuboike et al., in prep.). Regions with strong oxygen emission are identified and further divided into Ne-rich (red), Ne-/Si-rich (orange), and Si-rich (yellow) components. White contours indicate the distribution of $^{44}$Ti, highlighting the North bright spot. {\it Right}: Representative X-ray spectra extracted from the three PCA-defined regions. The spectra show a systematic progression from Ne-rich/Si-poor (red) to Si-rich/Ne-poor (yellow) compositions while maintaining strong oxygen emission, demonstrating that the PCA effectively captures compositional variations within the oxygen-rich ejecta. The $^{44}$Ti North bright spot broadly follows the distribution of oxygen-rich material but shows a stronger spatial association with the Ne-rich and Ne-/Si-rich components than with the most Si-rich regions.
}
\label{fig:pca}
\end{figure*}

\section{Observations and Data reduction}
In this study, we investigate the spatial distributions and spectral properties of the ejecta elements using X-ray observations (see section~\ref{dist}). For the analysis of the spatial distributions, we primarily utilize the 2004 Chandra observations. Cas A has been observed several times since launch with Chandra's Advanced CCD Imaging Spectrometer (ACIS-S) X-ray imager \citep[e.g.,][]{2000ApJ...528L.109H,2004ApJ...615L.117H,2012ApJ...746..130H}. We used the deepest ACIS-S data targeting Cassiopeia A observed in 2004, with a total exposure of about 1 Ms. We reprocessed the event files (from level 1 to level 2) to remove pixel randomization and to correct for charge-coupled device (CCD) charge-transfer inefficiences using CIAO version 4.17 and CalDB 4.12.2. The bad grades were filtered out and good time intervals were accepted. In addition, we use the Spitzer [Si~II] and NuSTAR $^{44}$Ti images of Cassiopeia A published by \citet{2009ApJ...693..713S} and \citet{2014Natur.506..339G}, respectively, for comparison with the X-ray data.

To investigate the detailed spectral structure of stable Ti, which is closely related to the production of $^{44}$Ti, we also use XRISM observations of Cas A (see section~\ref{ti44mass}). The XRISM observations were conducted during the XRISM commissioning phase and consisted of two pointings toward the southeastern (SE) and northwestern (NW) regions of the remnant \citep{2025PASJ...77S.171P}. In this study, we use the data from the SE pointing, where signatures of Ti were detected \citep{2026ApJ..1001...46S}. Data reduction utilized calibration files from the HEASARC Calibration Database (CALDB). Cleaned event lists were produced with HEASARC software v6.34, applying standard screening during post‐pipeline processing, resulting in a clean exposure time of 181.3 ks for the SE region. For spectral analysis, we selected only the highest‐resolution (``Hp'') primary events. The redistribution matrix file (RMF) was generated in “extra‐large” mode via \texttt{rslmkrmf}, and the ancillary response file (ARF) was generated with \texttt{xaarfgen}, adopting Cas A’s surface brightness profile as measured from a 2.0--8.0 keV Chandra X-ray image. The Resolve spectrum shown in this paper was taken from the Fe-rich region defined in \cite{2026ApJ..1001...46S}. The spectral fit was performed with XSPEC v12.14.1 (AtomDB v3.1.3) \citep{arnaud1996}, using the maximum-likelihood \textit{C}-statistic \citep{cash1979} over the 1.6--12.0 keV band.

\section{Ejecta distribution in Cas A}\label{dist}

\begin{figure*}[t]
 \begin{center}
  \includegraphics[bb=0 0 2336 663,width=18cm]{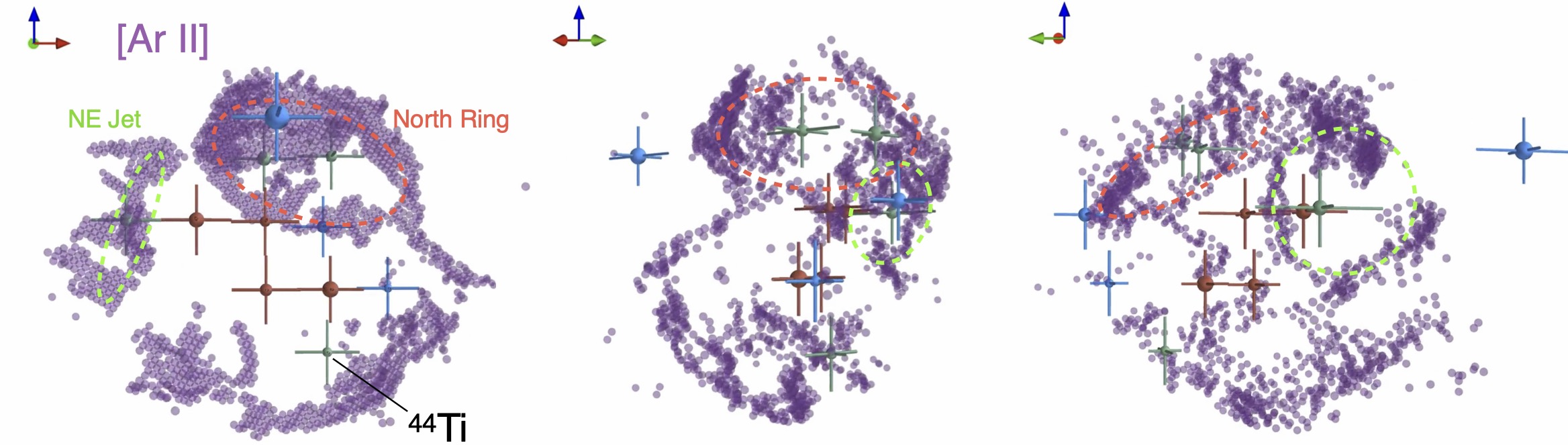}
 \end{center}
\caption{The 3D distribution of the observed $^{44}$Ti ejecta compared with the [Ar II] emission observed by Spitzer \citep{2010ApJ...725.2038D}. The $^{44}$Ti points are colored according to their positions relative to the reverse shock, following \cite{2017ApJ...834...19G}: red points lie inside the reverse shock, green points are near the reverse-shock radius, and blue points lie outside the reverse shock. This comparison of the 3D distributions was performed in Figure 10 of \cite{2017ApJ...834...19G}. The images from left to right present projections from different viewing angles, demonstrating that the intermediate-velocity $^{44}$Ti ejecta are contained within the NE jet and the North ring-like structure.}
\label{fig:f2}
\end{figure*}

Figure~\ref{fig:f1} presents a multiwavelength view of the ejecta distribution in Cassiopeia A, comparing the spatial distribution of radioactive $^{44}$Ti observed with NuSTAR to that of shocked and unshocked ejecta traced at X-ray and infrared wavelengths. These comparisons reveal that the relationship between $^{44}$Ti and other nucleosynthetic products is more complex than expected from a simple explosive nucleosynthesis scenario.

In the left panel of Figure~\ref{fig:f1}, the distribution of $^{44}$Ti is compared with shocked O-rich and Fe-rich ejecta observed in X-rays. A notable feature is the weak spatial correlation between the X-ray–bright shocked ejecta and the $^{44}$Ti emission. In particular, Fe, which is synthesized together with $^{44}$Ti during $\alpha$-rich freezeout in complete explosive Si burning, shows a distribution that largely avoids the central concentration of $^{44}$Ti. This behavior contrasts with expectations from standard explosion models and highlights a long-standing puzzle regarding the origin of $^{44}$Ti in Cas A. The shocked O-rich ejecta also exhibit a distribution that is mostly distinct from that of $^{44}$Ti, although localized overlaps are present. In particular, a bright O-rich region in the northeastern part of the remnant spatially coincides with enhanced $^{44}$Ti emission, and at the base of the NE jet the O-rich ejecta appear connected to the $^{44}$Ti distribution.

The right panel of Figure~\ref{fig:f1} compares the $^{44}$Ti distribution with unshocked Si-rich ejecta observed by Spitzer in the infrared \citep[e.g.,][]{2008ApJ...673..271R,2009ApJ...693..713S}. Unlike the shocked X-ray–emitting material, the unshocked Si is concentrated toward the interior of the remnant, where the reverse shock has not yet penetrated. Strikingly, the unshocked Si-rich ejecta exhibit a spatial distribution very similar to that of $^{44}$Ti, including a strong central concentration. This correspondence is consistent with recent studies of the inner ejecta in Cas A. In particular, \cite{2020ApJ...904..115L} argued that $0.47^{+0.47}_{-0.23}\ M_\odot$ of ejecta remains unshocked in the interior, with O and Si constituting the dominant components and no Fe detected. A comparable unshocked ejecta mass of $0.5$--$0.8\ M_\odot$ was independently derived by \cite{2022MNRAS.509.3163P}. The close spatial association between $^{44}$Ti and the unshocked Si-rich ejecta therefore suggests that a substantial fraction of the observed $^{44}$Ti resides in this cold, interior component.

\cite{2025ApJ...990..103S} suggested that an incomplete shell merger occurred in the progenitor of Cas A, resulting in the coexistence of Si-rich and Ne-rich oxygen ejecta within the remnant. If a significant fraction of $^{44}$Ti was synthesized in the oxygen layer affected by such a merger, its spatial distribution might be expected to correlate with Si-rich oxygen ejecta, which trace oxygen-burning products. Motivated by this expectation, we classify the chemical properties of oxygen-rich ejecta in Cas A and compare them with the distribution of $^{44}$Ti. Figure~\ref{fig:pca} presents this analysis, in which principal component analysis (PCA) is used to extract compositional variations within oxygen-dominated regions (Y. Kuboike et al., in prep.). In the left panel of Figure~\ref{fig:pca}, regions with strong oxygen emission are identified and further divided into Ne-rich (red), Ne-/Si-rich (orange), and Si-rich (yellow) components. Representative spectra extracted from these three regions are shown in the right panel. The spectral differences confirm that the PCA successfully captures a compositional transition from Ne-rich/Si-poor to Si-rich/Ne-poor ejecta. We find that the $^{44}$Ti ``North bright spot'' broadly follows the distribution of oxygen-rich ejecta; however, its spatial correlation appears stronger with the Ne-rich or Ne-/Si-rich components than with the most Si-rich regions. At the same time, the spectra indicate that even the regions classified as Ne-rich or Ne-/Si-rich contain a non-negligible amount of Si, suggesting that oxygen-burning products are present across these components to varying degrees. Thus, at present, it is difficult to conclude that the $^{44}$Ti observed in the North bright spot is of purely stellar origin. The PCA employed here is also useful for characterizing the properties of the shell merger in Cas A. A detailed spectral analysis and comparisons with theoretical models based on these results will be presented in a forthcoming paper (Y. Kuboike et al. 2026, in preparation).

Additional insight is provided by the three-dimensional comparison shown in Figure~\ref{fig:f2}, which juxtaposes the reconstructed 3D distribution of $^{44}$Ti with the [Ar II] emission observed by Spitzer \citep{2010ApJ...725.2038D}. This comparison follows the analysis presented in \cite{2017ApJ...834...19G} and highlights the spatial relationship between $^{44}$Ti and Ar-rich structures in Cas A. The intermediate-velocity $^{44}$Ti ejecta, shown in green, are found to be embedded within the base of the NE jet and within the ring-like structure in the northern region, both of which are prominent features in the [Ar II] emission. The spatial coexistence of $^{44}$Ti and Ar in these regions is particularly noteworthy, as Ar is one of the key seed nuclei involved in the synthesis of $^{44}$Ti in O–C shell merger environments. Their co-location therefore suggests a physical connection between the observed $^{44}$Ti and O-layer material enriched during late-stage stellar evolution. We note, however, that the elemental Ar abundance does not uniquely trace the abundance of the seed isotope $^{38}$Ar, as isotopic ratios can be modified by mixing conditions during shell mergers \citep{2026ApJ...997...41I,2026ApJ...997..314I}.

In contrast, the low-velocity $^{44}$Ti ejecta shown in red are confined to the interior of the remnant and are cleanly enclosed by the surrounding ejecta structure. Their location well inside the reverse shock strongly suggests an association with unshocked ejecta. The confinement of these $^{44}$Ti components within the remnant interior, together with their spatial coincidence with unshocked Si-rich material, further supports the interpretation that part of the $^{44}$Ti is linked to the cold inner ejecta rather than to shocked Fe-rich material.

\begin{figure}[t]
 \begin{center}
  \includegraphics[bb=0 0 939 894,width=8cm]{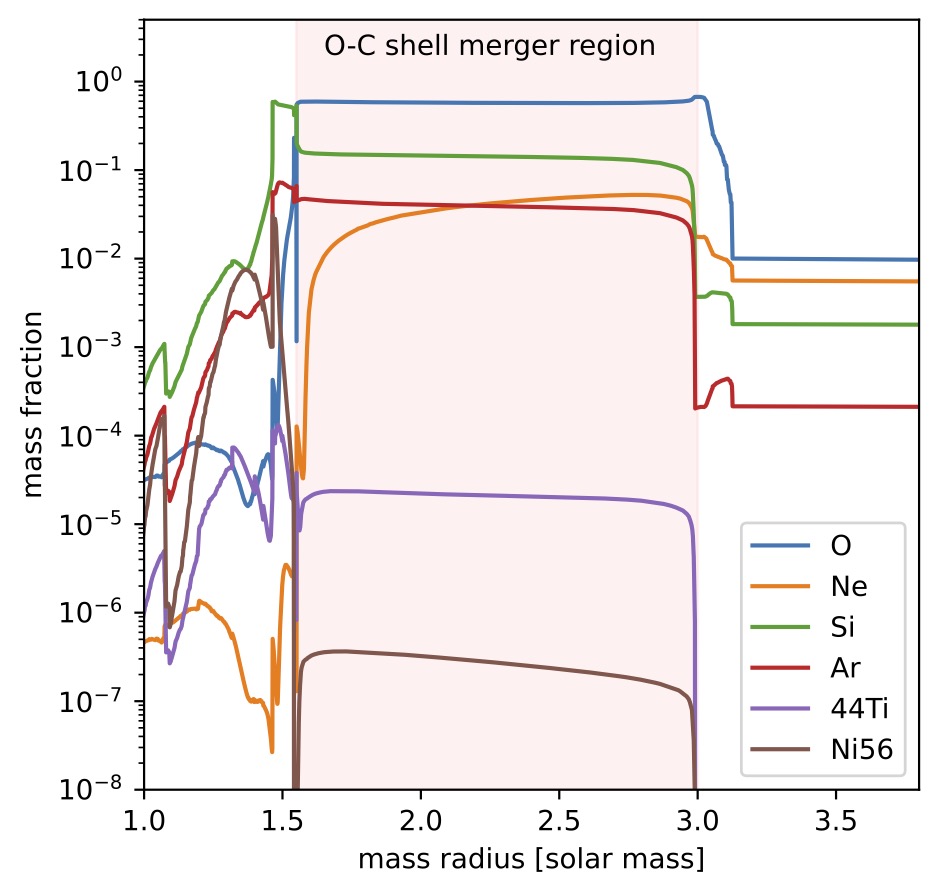}
 \end{center}
\caption{Mass fraction profile of the 15 $M_\odot$ Z = 0.02 model (not post-processed) from \cite{2018MNRAS.474L...1R} at the final time step before explosion. The O-C shell merger region is shaded to indicate its extent.}
\label{fig:f3}
\end{figure}

\begin{figure*}[t]
 \begin{center}
  \includegraphics[bb= 0 0 2340 1250,width=16cm]{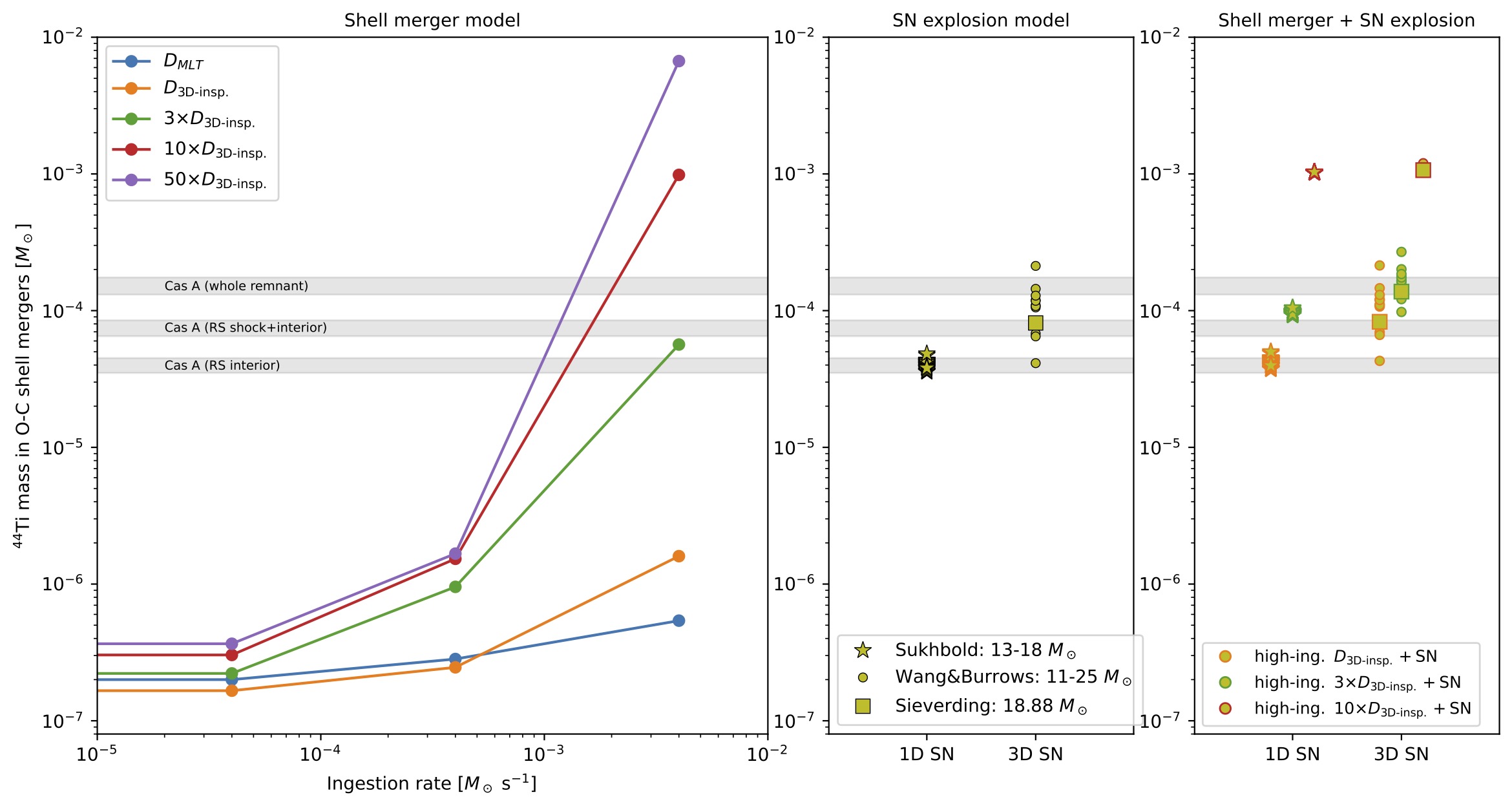}
 \end{center}
\caption{{\it Left}: Comparison between the $^{44}$Ti mass inferred from observations in Cas A and theoretical pre-supernova models by \cite{2026ApJ..1004...52I}. The horizontal axis shows the ingestion rate, while different colors indicate different mixing profiles. Gray horizontal bands indicate the ranges of the total $^{44}$Ti mass inferred for the whole remnant, the reverse-shock (RS) shock position + interior region, and the RS interior region from previous studies \citep{2017ApJ...834...19G}. {\it Middle}: Comparison of the observed $^{44}$Ti mass with predictions from supernova explosion models. Symbols represent predictions from one-dimensional models by \citet{2016ApJ...821...38S} and three-dimensional models by \citet{2023ApJ...957L..25S} and \citet{2024ApJ...974...39W}. {\it Right}: Same as the middle panel, but with the pre-supernova $^{44}$Ti contribution from the highest-ingestion-rate shell-merger models added to the supernova yields. The edge colors indicate the adopted shell-merger models, corresponding to $D_{\rm 3D\mbox{-}insp.}$, $3\times D_{\rm 3D\mbox{-}insp.}$, and $10\times D_{\rm 3D\mbox{-}insp.}$, as shown in the left panel. The gray horizontal bands in the middle and right panels are the same as those shown in the left panel.
}
\label{fig:f4}
\end{figure*}

Figure~\ref{fig:f3} provides a theoretical context for these observational results by showing the mass fraction profiles of a 15 $M_\odot$ pre-supernova model that undergoes an oxygen–carbon (O–C) shell merger shortly before core collapse. In this model, the shell merger substantially alters the chemical composition of the oxygen layer, producing an O-rich region that is enhanced in Si and depleted in Ne relative to standard progenitors. This composition closely resembles the Ne-poor and Si-rich oxygen-layer abundances recently identified in Cas A \citep{2025ApJ...984..185S,2025ApJ...990..103S}, suggesting that a similar shell merger occurred in its progenitor.

An important consequence of the O--C shell merger is its distinctive nucleosynthetic signature. The ingestion of C-shell material into the hot O-burning region supplies additional seed nuclei, including $^{43}$Ca, while vigorous convective mixing transports them to regions where further processing toward $^{44}$Ti becomes efficient.
In this environment, $^{44}$Ti can be synthesized through reaction pathways involving Ar and Ca isotopes, such as
$^{38}\mathrm{Ar}(\alpha,\gamma)^{42}\mathrm{Ca}(n,\gamma)^{43}\mathrm{Ca}(p,\gamma)^{44}\mathrm{Sc}(p,n)^{44}\mathrm{Ti}$ \citep{2026ApJ..1004...52I}. The characteristic temperatures at the bottom of the merged shell, around $\approx 2.8$ GK, are significantly lower than those required for complete explosive Si burning and do not favor the copious production of $^{56}$Ni. As a result, the merged oxygen layer is predicted to exhibit a high $^{44}$Ti-to-Fe ratio together with enhanced O and Si abundances.

Taken together, the weak spatial correlation between $^{44}$Ti and shocked Fe-rich ejecta, the close association of $^{44}$Ti with unshocked Si-rich material in the remnant interior, and the three-dimensional coexistence of intermediate-velocity $^{44}$Ti with Ar-rich structures would suggest that the ejecta distribution in Cas A retains signatures of late-stage stellar evolution. These observational characteristics are naturally explained if at least part of the observed $^{44}$Ti originated in the oxygen layer prior to explosion, potentially in connection with an O–C shell merger in the progenitor star.

\section{Mass of Pre-supernova $^{44}$Ti}\label{ti44mass}

In this section we estimate an observational upper limit on the amount of $^{44}$Ti in Cassiopeia A that could plausibly originate from pre-supernova nucleosynthesis, motivated by the ejecta distribution discussed in Section 2. The key empirical clue is that the low- and intermediate-velocity components of $^{44}$Ti, corresponding to the red and green points in Figure~\ref{fig:f2}, are more closely associated with O-layer tracers (O, Si, and Ar) than with Fe-rich shocked ejecta. If we adopt an extreme assumption that all of these low- and intermediate-velocity components were synthesized in the oxygen layer during a pre-collapse O–C shell merger, then the implied pre-supernova contribution is $\sim 7\times10^{-5}\ M_\odot$ (roughly half of the total $^{44}$Ti mass) inferred for Cas A. We emphasize that this value should be regarded as an upper limit, not a preferred estimate, because it attributes essentially all O-layer–associated $^{44}$Ti to the pre-supernova channel. Although derived using a different approach, this upper limit is
consistent with \citet{2026ApJ..1003L...2B}, who estimated that
C--O shell mergers could contribute up to $\sim 20$--$30\%$ of
the total $^{44}$Ti in Cas~A. Their estimate lies within the
observational upper limit derived here.

To support the $^{44}$Ti distribution concentrated in the unshocked regions revealed by NuSTAR, we also considered the possibility of detecting X-ray emission lines associated with the decay of Ti via orbital electron capture, which would produce K-shell fluorescence lines of Sc at around 4 keV \citep{2010ApJ...724L.161B,2025ApJ...989..199H,2025A&A...700A.254G}. However, as discussed in Appendix~\ref{ap:44sc} and Figure~\ref{fig:44sc}, the expected line flux is low and severe line confusion with much stronger thermal Ca emission, together with current instrumental limitations, makes a secure detection challenging with existing data.

Figure~\ref{fig:f4} (left) compares the upper limit of $^{44}$Ti with the O–C shell-merger nucleosynthesis models of \citep{2026ApJ..1004...52I}. The comparison indicates that the observationally inferred pre-supernova contribution can be approached for ingestion rates of roughly $\gtrsim 10^{-3}~M_\odot~{\rm s^{-1}}$ and diffusion coefficients of $\gtrsim 3D_{\rm 3D\mbox{-}insp.}$, corresponding to more vigorous convective mixing than in standard one-dimensional stellar-evolution calculations represented by $D_{\rm MLT}$. In particular, models adopting convective velocities several times larger than the nominal one-dimensional values, as suggested by three-dimensional hydrodynamic simulations \citep{2006ApJ...637L..53M,2007ApJ...667..448M,2017MNRAS.465.2991J,2020MNRAS.491..972A,2020ApJ...890...94Y,2024MNRAS.533..687R}, can reach the $^{44}$Ti mass implied by the red+green components. 

The middle and right panels of Figure~\ref{fig:f4} illustrate how the pre-supernova contribution may complement explosive $^{44}$Ti production. The middle panel shows the $^{44}$Ti yields predicted by one- and three-dimensional supernova explosion models, while the right panel shows the total obtained by adding the yields from the highest-ingestion-rate shell-merger models to the supernova yields. Under this simple assumption, even some one-dimensional explosion models can reach the $^{44}$Ti mass inferred for Cas A when combined with a pre-supernova contribution, while the three-dimensional models span a broader range of solutions consistent with the observations. We emphasize, however, that this comparison is illustrative: simply adding the pre-supernova and explosive yields does not account for possible destruction, reprocessing, or fallback of the $^{44}$Ti synthesized before core collapse.

At the same time, it is unlikely that essentially all of the $^{44}$Ti in Cas A is of O-layer origin. First, the spatial correspondence between shocked O-rich ejecta and $^{44}$Ti is suggestive but not one-to-one, and substantial fractions of the shocked O emission do not coincide with $^{44}$Ti. Second, the mass of unshocked ejecta inferred for the remnant interior, $\sim 0.5$--$0.8 ~M_\odot$, represents only a subset of the total ejecta mass of Cas A, which is often estimated to be a few solar masses. If the $\sim 4\times10^{-5} ~M_\odot$ of $^{44}$Ti associated with the unshocked interior component were to be produced entirely in the pre-supernova oxygen layer, the required production efficiency would need to be correspondingly high. Finally, if the full $^{44}$Ti inventory were attributed to the stellar-evolution channel, an additional question arises: where is the $^{44}$Ti component that should accompany Fe-group material synthesized in alpha-rich freeze-out during the explosion? The presence of Fe thought to originate from the deepest, hottest ejecta implies that at least some fraction of $^{44}$Ti should also be produced explosively, even if part of the observed distribution reflects pre-supernova processing. 

Recent three-dimensional core-collapse supernova simulations predict substantial explosive production of $^{44}$Ti and demonstrate that its yield can vary significantly with the explosion dynamics \citep[e.g.,][]{2023ApJ...957L..25S,2024ApJ...974...39W}. Moreover, the nucleosynthetic outcome may depend sensitively on the electron fraction, and $^{44}$Ti production can be suppressed in proton-rich ejecta affected by neutrino interactions \citep[e.g.,][]{2018ApJ...852...40W}. These results caution against interpreting the observed $^{44}$Ti distribution in Cas A as a unique signature of pre-supernova nucleosynthesis. Together with the comparison shown in Figure~\ref{fig:f4}, these results favor a scenario in which both pre-supernova and explosive nucleosynthesis contribute to the total $^{44}$Ti budget. A robust decomposition of the $^{44}$Ti budget likely requires next-generation simulations that combine realistic late-stage progenitor evolution, including shell mergers and convective-reactive ingestion, with multi-dimensional neutrino-driven explosion dynamics and detailed nucleosynthesis. Such integrated models will be essential for determining how much of the $^{44}$Ti in Cas A is inherited from pre-supernova stellar evolution and how much is synthesized during the explosion.

\begin{figure}[t]
 \begin{center}
  \includegraphics[bb=0 0 1200 1000,width=8cm]{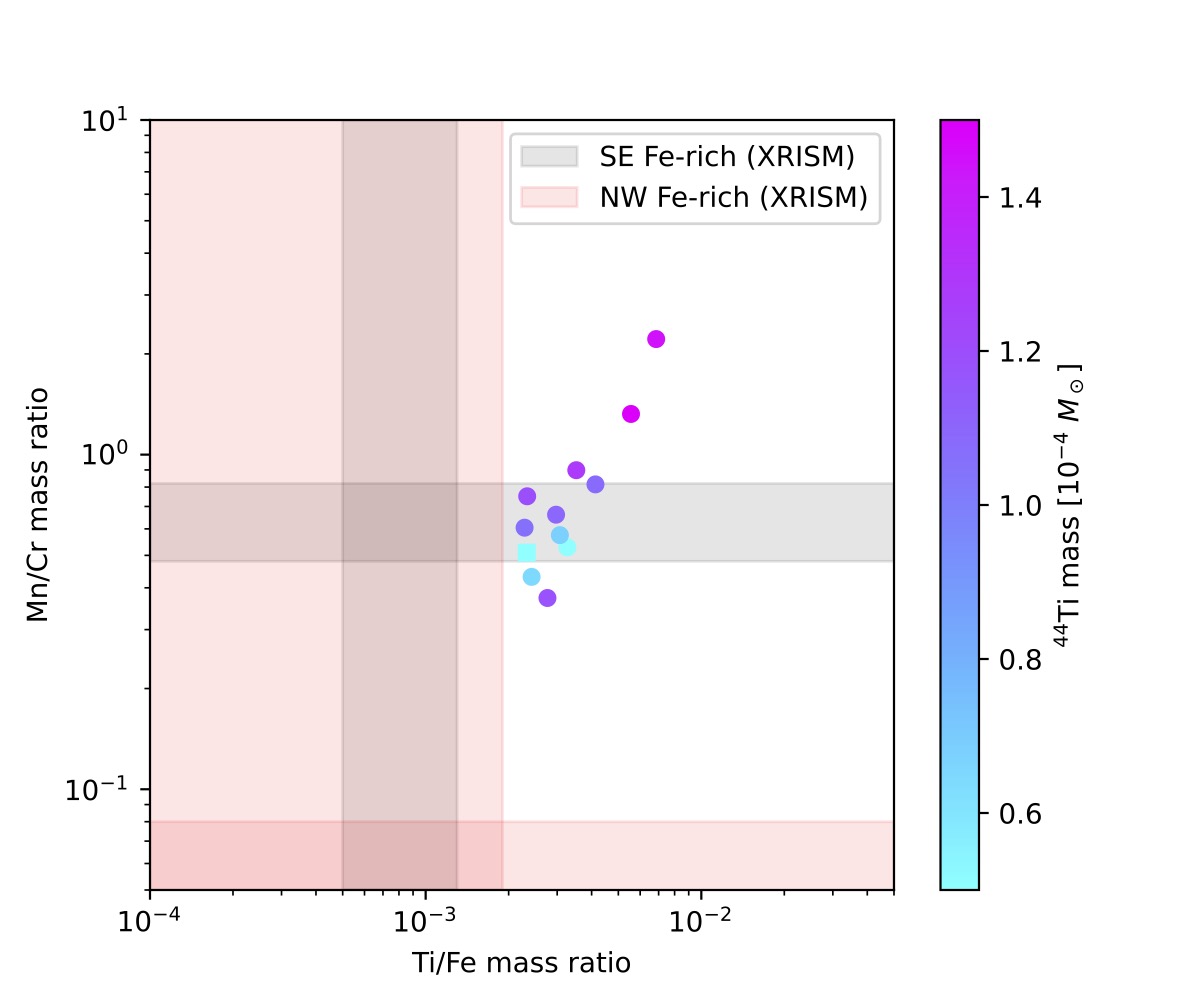}
 \end{center}
\caption{Stable Ti/Fe versus Mn/Cr mass ratios for the three-dimensional core-collapse supernova models of \citet{2024ApJ...962...71W}. Each point represents an individual explosion model, with the color indicating the corresponding radioactive $^{44}$Ti yield \citep{2024ApJ...974...39W}. The gray and red shaded regions show the abundance ratios measured in the southeastern (SE) and northwestern (NW) Fe-rich ejecta of Cas A with XRISM/Resolve \citep{2026ApJ..1001...46S}, respectively. The comparison illustrates the relationship between the production of radioactive $^{44}$Ti and stable Fe-group elements in the supernova ejecta.
}
\label{fig:mncr_tife}
\end{figure}

\begin{figure*}[t]
 \begin{center}
  \includegraphics[bb=0 0 1858 1111,width=16cm]{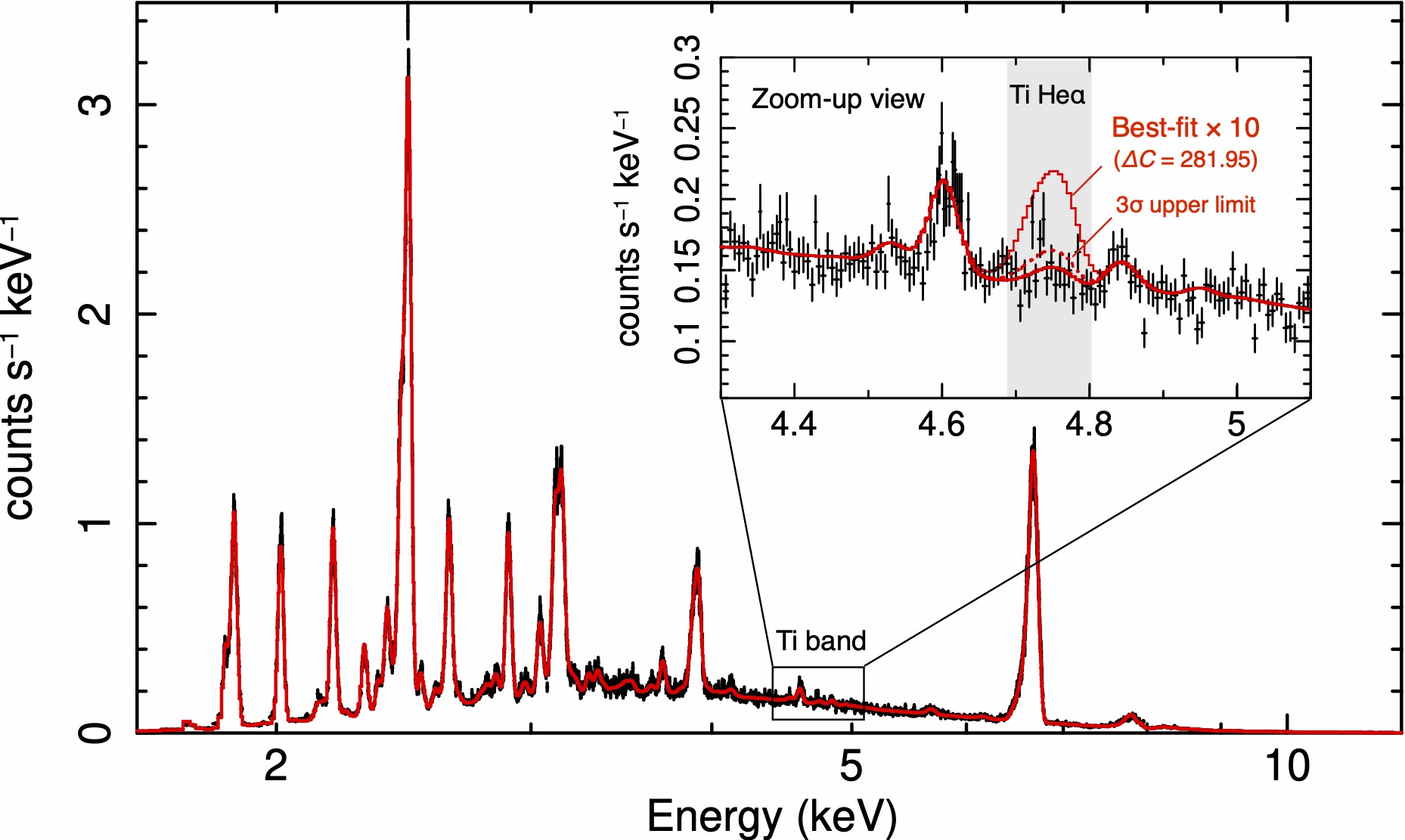}
 \end{center}
\caption{XRISM/Resolve spectrum (1.6--12 keV) extracted from the Fe-rich region in the SE pointing of the Cas A observations. The black data show the observed spectrum, while the red curve indicates the best-fit plasma emission model shown in \cite{2026ApJ..1001...46S}. The inset panel presents a zoomed-in view of the Ti He$\alpha$ band, where the best-fit model (red) is compared with a model in which the Ti abundance is artificially increased by a factor of 10. The latter is intended to mimic the Ti/Fe mass ratio of $\approx 10^{-2}$ expected in models producing a large amount of $^{44}$Ti (see Figure~\ref{fig:mncr_tife}). The difference between these two models corresponds to $\Delta C = 281.95$, which is unlikely to arise from statistical fluctuations alone (corresponding to a p-value of $\approx 10^{-63}$). The red dashed curve indicates the 3$\sigma$ upper limit on the Ti line emission ($\Delta C = 9.0$), corresponding to a Ti/Fe mass ratio of 0.0024.}
\label{fig:f6}
\end{figure*}

\section{Explosive $^{44}$Ti production and stable Fe-group abundance ratios}

The abundances of stable Fe-group elements in the Fe-rich ejecta of Cas~A provide a useful probe of the physical conditions in the innermost regions of the explosion. In particular, the production of Ti, Cr, and Mn is sensitive to the thermodynamic conditions, electron fraction, and neutrino processing of the ejecta \citep{2021Natur.592..537S,2023ApJ...954..112S,2026ApJ..1001...46S}. These same conditions are also expected to play an important role in the explosive production of $^{44}$Ti. Stable abundance ratios such as Ti/Fe and Mn/Cr in the Fe-rich ejecta may therefore provide an independent constraint on explosion models that produce large amounts of $^{44}$Ti.

Motivated by this connection, we compare the XRISM measurements with the stable Fe-group yields of \citet{2024ApJ...962...71W}, combined with the corresponding radioactive $^{44}$Ti yields presented in \citet{2024ApJ...974...39W}. Here, we estimated the synthesized masses of the stable elements from the published isotopic-yield tables of \citet{2024ApJ...962...71W} by summing the abundances of isotopes that are either stable or decay into the corresponding stable species within 350 years after the explosion, neglecting isotopes with yields below $10^{-8}\ M_\odot$. Figure \ref{fig:mncr_tife} shows the predicted Mn/Cr and Ti/Fe mass ratios, with the color of each model indicating its total radioactive $^{44}$Ti yield. The models exhibit a systematic tendency for both Mn/Cr and Ti/Fe to increase toward models with larger $^{44}$Ti yields. This behavior is qualitatively consistent with the sensitivity of Fe-group nucleosynthesis to the thermodynamic conditions and electron fraction of the innermost ejecta. In particular, multi-dimensional explosion calculations show that proton-rich ejecta produced by neutrino interactions can substantially alter the yields of Ti and neighboring Fe-group nuclei \citep{2024ApJ...962...71W}.

A notable tension emerges when these predictions are compared with the Fe-rich ejecta observed by XRISM. The three-dimensional models producing the largest $^{44}$Ti masses tend to occupy regions of the Mn/Cr–Ti/Fe plane with abundance ratios higher than those inferred for the southeastern and northwestern Fe-rich ejecta. Thus, the large $^{44}$Ti yields in these long-term 3D models are closely associated with extended, predominantly proton-rich neutrino-driven winds, in which $^{44}$Ti continues to be produced during the late-time freeze-out as the ejecta cool \citep{2024ApJ...974...39W}. Such strongly neutrino-processed ejecta can also alter the production of stable Fe-group nuclei, providing a natural motivation for comparing the predicted Ti/Fe and Mn/Cr ratios with those measured in the Fe-rich ejecta of Cas~A. The Fe-rich ejecta therefore provide an important constraint on models in which most or all of the observed $^{44}$Ti is synthesized through strongly neutrino-processed explosive nucleosynthesis.

This tension can also be tested directly using the XRISM/Resolve spectrum of the SE Fe-rich region in CasA. Figure \ref{fig:f6} compares the best-fit plasma model with an otherwise identical model in which the Ti abundance is artificially increased to match Ti/Fe $\sim10^{-2}$, representative of the values predicted by high-$^{44}$Ti explosion models. The Ti-enhanced model is strongly rejected by the XRISM spectrum, yielding a difference of $\Delta C = 281.95$, which is extremely unlikely to arise from statistical fluctuations alone (corresponding to a p-value of $\approx10^{-63}$). In addition, the XRISM spectrum places a 3$\sigma$ upper limit of 0.0024 on the Ti/Fe mass ratio in the SE Fe-rich ejecta ($\Delta C=9.0$). This limit lies below the Ti/Fe ratios predicted by most of the current three-dimensional explosion models shown in Figure~\ref{fig:f6}, further highlighting the difficulty of simultaneously reproducing the large $^{44}$Ti yield and the stable Fe-group composition observed in Cas A. This provides a direct observational demonstration that the large stable Ti/Fe ratios associated with some high-$^{44}$Ti explosion models are inconsistent with the SE Fe-rich ejecta.

We caution that this comparison does not by itself exclude an explosion-only origin of $^{44}$Ti. The XRISM measurements sample particular shocked Fe-rich structures, whereas the observed radioactive $^{44}$Ti mass represents the remnant as a whole, and substantial spatial variations in the Fe-peak composition are already evident within Cas~A \citep{2026ApJ..1001...46S}. Moreover, the detailed abundance ratios depend sensitively on the local thermodynamic and electron-fraction histories of individual ejecta parcels. Nevertheless, the apparent tendency for high-$^{44}$Ti models to predict elevated Mn/Cr and Ti/Fe suggests that simultaneously reproducing the large radioactive $^{44}$Ti yield and the stable Fe-group composition of the Fe-rich ejecta may be non-trivial for models relying on very strong neutrino processing alone.

In this context, a pre-supernova contribution from O–C shell-merger nucleosynthesis provides a possible way to reduce the amount of $^{44}$Ti that must be synthesized in the deepest explosive ejecta. Such a contribution would relax the requirement for the explosion itself to produce the entire observed $^{44}$Ti, while still allowing a substantial fraction of $^{44}$Ti to originate from high-entropy, neutrino-processed ejecta. The combination of radioactive $^{44}$Ti measurements and stable Fe-group abundance ratios therefore helps constrain the relative contributions of pre-supernova and explosive nucleosynthesis in Cas~A.

\section{Summary and Conclusions}

We have investigated the origin of $^{44}$Ti in the Cassiopeia A supernova remnant by combining its spatial and kinematic distribution with multiwavelength diagnostics of the ejecta composition. Our analysis shows that the ejecta distribution of $^{44}$Ti cannot be fully understood by considering explosive nucleosynthesis alone, nor does it require that all of the observed $^{44}$Ti be inherited from pre-supernova stellar evolution.

The low- and intermediate-velocity components of $^{44}$Ti exhibit strong spatial associations with O-layer material, including unshocked Si-rich ejecta in the remnant interior and Ar-rich structures associated with the NE jet and the northern ring. These correlations naturally arise if part of the $^{44}$Ti was synthesized in the oxygen layer prior to explosion, potentially during an O–C shell merger in the progenitor star. Based on these associations, we estimate an observational upper limit of $\sim 7\times10^{-5}\ M_\odot$ for the amount of $^{44}$Ti that could originate from pre-supernova nucleosynthesis.

Comparison with recent O–C shell merger models demonstrates that reaching this upper limit requires high ingestion rates and convective velocities larger than those typically found in current one-dimensional stellar evolution simulations. At the same time, several lines of evidence argue against a purely pre-supernova origin for all of the $^{44}$Ti in Cassiopeia A. The weak spatial correspondence between shocked O-rich ejecta and $^{44}$Ti, the limited mass of unshocked interior ejecta relative to the total ejecta mass, and the presence of Fe-rich material thought to originate from $\alpha$-rich freeze-out all point to a significant explosive contribution.

Recent three-dimensional core-collapse supernova simulations further show that explosion physics alone can reproduce a wide range of $^{44}$Ti/Fe ratios and morphologies consistent with Cassiopeia A, highlighting the importance of neutrino-driven processes and electron-fraction effects. At the same time, the stable Fe-group composition of the Fe-rich ejecta provides an independent constraint on explosive $^{44}$Ti production. Three-dimensional explosion models producing the largest $^{44}$Ti yields tend to predict elevated Ti/Fe and Mn/Cr ratios compared with those measured by XRISM, and the direct spectral constraint on the Ti abundance further highlights this tension. These results suggest that reproducing the total $^{44}$Ti mass of Cas~A through strongly neutrino-processed explosive nucleosynthesis alone may be challenging. A pre-supernova contribution would reduce the amount of $^{44}$Ti that must be synthesized in the deepest explosive ejecta while still allowing a substantial explosive component. Taken together, our results support a mixed-origin scenario in which the observed $^{44}$Ti reflects contributions from both pre-supernova stellar evolution and explosive nucleosynthesis.

Future progress will require next-generation simulations that self-consistently couple late-stage progenitor evolution, including shell mergers and convective-reactive ingestion, with multi-dimensional explosion dynamics and detailed nucleosynthesis. Such models, together with improved spatially resolved observations, will be essential for quantitatively disentangling the relative contributions of pre-supernova and explosive processes to the $^{44}$Ti budget in Cas A.

\vspace{0.5cm}
\noindent 
{\bf Acknowledgments}: We are grateful to Dr. Brian Grefenstette for providing the $^{44}$Ti observational results and for his helpful comments. This work was partly supported by Japan Society for the Promotion of Science Grants-in-Aid for Scientific Research (KAKENHI) Grant Number P23K13128 (TS) and 24KJ1485 (KM). KM was supported by the JSPS Core-to-Core Program (grant number JPJSCCA20220002) for a research visit to the University of Victoria. JML was supported by basic research funds of the Office of Naval Research. We acknowledge the use of ChatGPT (OpenAI) for assistance with English-language editing and improving the readability of the manuscript. The authors reviewed and take full responsibility for the final text.

\appendix

\section{XRISM/Resolve view of $^{44}$Sc line in Cassiopeia A}\label{ap:44sc}

\begin{figure*}[t]
 \begin{center}
  \includegraphics[bb=0 0 2100 900,width=18cm]{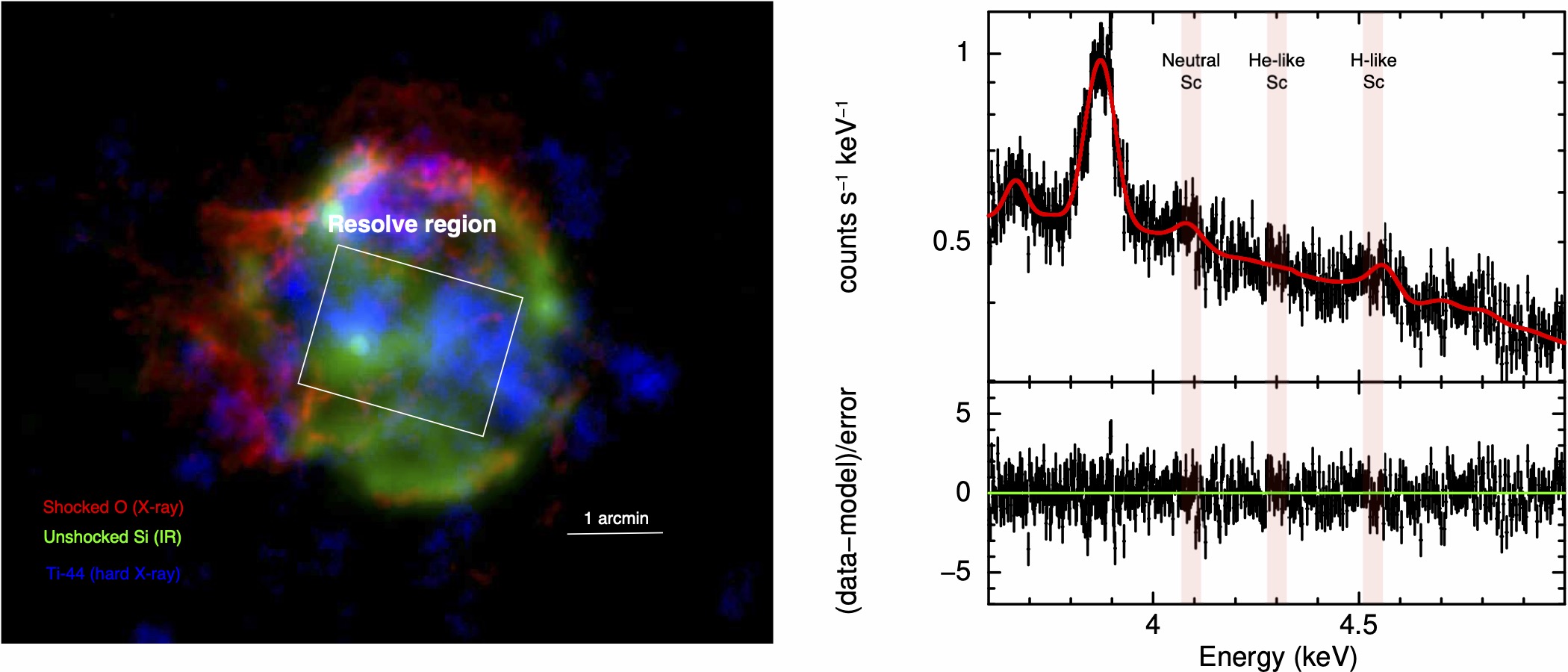}
 \end{center}
\caption{{\it Left}: Three-color image of Cassiopeia A, showing shocked oxygen (red), unshocked silicon (green), and $^{44}$Ti (blue). The overlaid polygon indicates the Resolve extraction region selected from the XRISM northwest pointing, chosen to closely match the projected distribution of $^{44}$Ti within the field of view.
{\it Right}: XRISM Resolve spectrum extracted from this region. Shaded regions mark the expected energies of Sc K$\alpha$ fluorescence lines produced by the decay of $^{44}$Ti under different post-decay electronic configurations.}
\label{fig:44sc}
\end{figure*}

The electron-capture decay of $^{44}$Ti produces K-shell vacancies in the daughter nucleus $^{44}$Sc, which may result in K$\alpha$ fluorescence emission at energies near 4.09~keV. Motivated by this process, we examined whether such Sc X-ray lines could be detectable in Cas~A with current X-ray observations. We used the XRISM Resolve NW pointing and selected pixel regions whose projected locations best trace the known distribution of $^{44}$Ti within the field of view (Figure~\ref{fig:44sc} left). From this Resolve region, we extracted a spectrum and fitted it using a two-temperature thermal plasma model together with an additional power-law component to account for non-thermal emission  (Figure~\ref{fig:44sc} right). As highlighted in the figure, the Sc emission lines following the decay of $^{44}$Ti are expected to appear at around 4.09 keV for neutral or weakly ionized $^{44}$Ti, at around 4.3 keV if $^{44}$Ti is near a Li-like ionization state, and at around 4.53 keV in the case of He-like $^{44}$Ti.

The situation is complicated by the presence of strong thermal Ca emission. In Cas~A, the Ca~XX Ly$\alpha$ line at 4.107~keV is significantly brighter than the expected Sc fluorescence signal. Plasma modeling in this energy band is therefore non-trivial: when multi-temperature plasma components are allowed, a wide variety of spectral models can reproduce the observed data. Robust discrimination between weak radioactive Sc features and thermal Ca emission would thus require high-resolution spectroscopy, such as that provided by XRISM. At present, however, limitations in photon statistics and the effects of line broadening prevent a definitive interpretation.


Following an order-of-magnitude estimate, an initial $^{44}$Ti mass of $10^{-4}\ M_\odot$ corresponds to $2.7 \times 10^{51}$ nuclei at the time of explosion. After $\sim 340$~yr, this implies $\sim 5 \times 10^{49}$ remaining $^{44}$Ti nuclei and a present-day decay rate of $\sim 1.9 \times 10^{40}$~s$^{-1}$. Assuming a K-shell fluorescence yield of order 0.1 photons per vacancy, the expected Sc K$\alpha$ flux at a distance of 3.4~kpc is $\sim 10^{-6}$~ph~cm$^{-2}$~s$^{-1}$. Assuming the Resolve effective area of 150 cm$^2$, line width of 100 eV, we expect 0.002 counts s$^{-1}$ keV$^{-1}$ for $^{44}$Sc line in the Resolve spectrum, however this implies that drawing a clear conclusion is difficult with the current situation.

Given these considerations, we conclude that, with current observations, the detection of $^{44}$Sc X-ray emission lines in Cas~A is highly challenging and cannot be claimed with confidence. Future high-resolution, high-throughput X-ray spectroscopy or XRISM deep observations of Cas A will be essential to revisit this issue.

\bibliography{sample701}{}
\bibliographystyle{aasjournalv7}



\end{document}